
\tolerance=5000
\input phyzzx
\nopubblock
\titlepage
\line{\hfill UFIFT-HEP-92-32}
\line{\hfill }
\line{\hfill }
\title{Third Quantization and Black Holes}
\author{Michael McGuigan\foot{Research supported in part by the U.S.
Department of
     Energy under contract no. DE-FG05-86ER-40272. } }
\vskip.2cm
\centerline{\it Institute for Fundamental Theory}
\centerline{\it University of Florida}
\centerline{\it Gainesville, Fl. 32611 }

\abstract{Recent results on the quantum mechanics of black holes indicate
that the spacetime singularity can be avoided if the space of fields
is extended in its domain as well as the vanishing of beta functions
required as an equation in target space. Such an approach seems to inevitably
lead to the quantization in this space that is the so-called third
quantization. We discuss some of the implications of third quantization
for the physics of blackholes in two dimensions. By discretizing
the transverse dimensions describing the horizon it may also be possible
to describe regulated four dimensional gravity in this manner and perhaps
shed some light on the meaning of black hole entropy.}
\endpage

\REF\dyso{F.Dyson, Institute for Advanced Study Report, 1975 (unpublished);
Y.Zeldovich, Sov.Phys.JETP45,9 (1977); S.Hawking, Phys.Rev.D37,904(1988);
S.Hawking and R.Laflamme, Phys.Lett.B209,39 (1988); S.Hawking,
Mod.Phys.Lett.A5,145,453 (1990)}
\REF\presk{J.Preskill (Caltech) CALT-68-1819, Jan.1992,
hep-th@xxx.lanl.gov-9209058}
\REF\bowi{M.J.Bowick, S.Giddings, J.Harvey, G.Horowitz and A.Strominger,
Phys.Rev.Lett.61,2823 (1988)}
\REF\giddi{S.Giddings and A.Strominger, Nucl.Phys.B306, 890 (1988)}
\REF\russ{J. G. Russo, L. Susskind and L. Thorlacius, "Black Hole
Evaporation in 1+1 dimensions," SU-ITP-92-4.}
\REF\suss{ L. Susskind and L. Thorlacius,
Nucl. Phys. B382 (1992) 123.}
\REF\cghs{C.Callan, S.Giddings, J.Harvey and A.Strominger, Phys.Rev.D45,1005
(1992)}
\REF\deal{S.P.DeAlwis, Phys.Lett.B289,278 (1992);
(Colorodo U.), Colo-hep-284-REV, Jun.1992'
hep-th@xxx.lanl.gov-99206020;
(Colorodo U.), Colo-hep-288-REV, Oct.1992,
hep-th@xxx.lanl.gov-99207095
}
\REF\bila{A.Bilal and C.Callan, (Princeton U.) PUPT-1320-REV Aug.1992,
hep-th@xxx.lanl.gov-99205089}
\REF\gidd{S.Giddings and A.Strominger, (UC, Santa Barbara)
UCSBTH-92-28, Jul.1992, hep-th@xxx.lanl.gov-99207034}
\REF\polc{J.Polchinski, Nucl.Phys.B346,253 (1990); S.Das and A.Jevicki,
Mod.Phys.Lett.A5,1639 (1990)}
\REF\mcg{M.McGuigan, Phys.Rev.D43,1199 (1991)}
\REF\betat{C.Callan and Z.Gan, Nucl.Phys.B272,647 (1986);
S.Das and B.Sathiapalan, Phys.Rev.Lett.56,2664 (1986)}
\REF\banks{T.Banks and J.Lykken, Nucl.Phys.B331,173 (1990)}
\REF\doro{A.G.Doroshkevich and I.D.Novikov, Sov.Phys.JETP47,1 (1978)}
\REF\chit{D.M.Chitre and J.B.Hartle, Phys.Rev.D16, 251 (1977)}
\REF\bro{J.D.Brown and E.A.Martinez, Phys.Rev.D42, 1931 (1990)}
\REF\moor{G.Moore and N.Seiberg, Nucl.Phys.B362,665 (1991)}
\REF\cham{A.H.Chamseddine, Phys.Lett.B256, 379 (1991)}
\REF\carl{S.Carlip, Phys.Rev.D42, 2647 (1990)}
\REF\witt{E.Witten, Nucl.Phys.B311, 46 (1988)}
\REF\dewi{B.S.DeWitt, Phys.Rev.160, 1113 (1967)}
\REF\birr{N.D.Birrel and P.C.W.Davies, "Quantum fields in curved space",
Cambridge University Press (1982)}
\REF\full{S.Fulling, Phys.Rev.D7,2850 (1973); R.Hagg, H.Narnhofer and
U.Stein, Comm.Math.Phys.94,219 (1984)}
\REF\verl{H.Verlinde and E.Verlinde, Nucl.Phys.B371,246 (1992)}
\REF\bard{W.Bardeen and R.Pearson, Phys.Rev.D14,547 (1976)}
\REF\griff{P.Griffin, Nucl.Phys.B372,270 (1992)}
\REF\pois{E.Poisson and W.Israel, Phys.Rev.D41,1796 (1990)}
\REF\frau{S.Frautchi, Phys.Rev.D3,2821 (1971); S.Carlitz,
Phys.Rev.D5,3231 (1972)}
\REF\ruba{V.Rubakov, Phys.Lett.B214,503 (1988)}
\REF\fisc{W.Fischler, I.Klebanov, J.Polchinski and L.Susskind,
Nucl.Phys.B327,157 (1989)}
\REF\gibb{G.Gibbons, Phys.Lett.A61,3 (1977);
G.Gibbons and S.Hawking, Phys.Rev.D15,2752 (1977)}

An interesting proposal for recovering the information lost in
blackhole evaporation involves the nucleation of a baby universe.
\refmark{\dyso}. In this picture the quantization of gravity permits
topology change
to take place during gravitational collapse and the new universe
created carries away all the information about the initial quantum state
(For a nice discussion of this and other issues involving the information loss
paradox see \refmark{\presk}).
Also systems such as the axionic blackhole \refmark{\bowi} which
are problematic in
the sense that the axionic quantum hair would be present after blackhole
evaporation also admit a wormhole solution which can carry off
the axionic quantum number \refmark{\giddi}. In any case
the non-renormalizablity of gravity inhibits the inclusion
of such processes in usual four dimensional general relativity.
Clearly it would useful
to study a simplified model where the impact of topology change
on blackhole dynamics can be adequately discussed.

Recently much attention has been focused on two dimensional dilaton
gravity as it is a renormalizable theory containing black hole
solutions. However results of these studies indicate that a semiclassical
approximation breaks down just before a naked singularity
appears \refmark{\russ,\suss}. Thus in order to address questions about the
last stages of Hawking evaporation as well as to incorporate
the effects of topology change on black hole radiance it
is necessary to go beyond the semiclassical approximation and
quantize the $2d$ dilaton gravity system. A useful approach to
quantization was taken by \refmark{\cghs} who chose the conformal gauge
and obtained as consistency conditions the vanishing of the beta
functions in target space \refmark{\deal, \bila, \gidd}.
Such an approach seems to inevitably
lead to the quantization in this space that is the so-called third
quantization. Indeed a closely related model
is the $c=1$ matrix model and there the target space was indeed
quantized with the ground state of the system being promoted to
the tachyon field \refmark{\polc}.

In this paper we discuss the $2d$ dilaton gravity arising from the reduction
of 3+1 dimensions to 1+1 dimensions upon the assumption of spherical
symmetry. This leads to a description of the s-wave sector of
general relativity as well as blackhole solutions with the usual
values of blackhole temperature and entropy. Further description
of the beta function technique applied to this and other
reductions of general relativity to 1+1 dimensions as well as
additional references are contained in
\refmark{\mcg}.

We start  in four dimensions
and make a spherical symmetric ansatz for the metric which gives
$$ [g_{\mu \nu}] = \pmatrix {g_{00} & g_{01} & 0 & 0\cr
                             g_{10} & g_{11} & 0 & 0\cr
                                0   &    0   & L^2e^{-2\phi} & 0\cr
                                0   &    0 & 0 & L^2e^{-2\phi}\sin^2{\theta} }
$$
The Einstein Hilbert action then takes the form
$$ S = {L^2 \over 4 G }
          \int d^2x \sqrt{-g} e^{-2 \phi} \left( R + 2 (\nabla \phi)^2
              + 2 L^{-2}e^{2 \phi} \right ) .       \eqn\1  $$
A useful gauge
choice for models of this type is the conformal gauge
$g_{ab} = e^{2\rho}\hat{g}_{ab}$. In this
gauge the action becomes:
$$ S = {L^2 \over 4 G }
          \int d^2x \sqrt{-\hat{g}} e^{-2 \phi} \left(
2\hat{g}^{ab}\partial_a\phi \partial_b\phi -
4\hat{g}^{ab}\partial_a\phi\partial_b\rho +
\hat{R}
              + 2 L^{-2}e^{2 \phi+2\rho} \right ) .         $$
This form of the action can be easily compared with the sigma model action
$$ S = -{1\over 4\pi \alpha '}
          \int d^2x \sqrt{-\hat{g}}  \left(
\hat{g}^{ab}\partial_aX^{I} \partial_bX^JG_{IJ}(X) + \alpha '\hat{R}\Phi(X)
              + T(X)  \right ) .       $$
with $G_{IJ}(X)$, $\Phi(X)$ and $T(X)$ given by
$$G_{IJ}(X)dX^IdX^J =
{\pi L^2\alpha '\over G}(-2d\phi d\phi + 4d\phi d\rho)e^{-2\phi}$$
$$\Phi(X) = -{\pi L^2 \over G}e^{-2\phi}$$
$$T(X)= -{2\pi \alpha ' \over G}e^{2\rho}$$
The metric is indeed flat as can be seen by the coordinate choice
$X^0= {\ell \over 2}(e^{-2\phi}+2\rho-\phi)$ and
$X^1= {\ell \over 2}(e^{-2\phi}-2\rho+\phi)$
where $\ell = \sqrt{\alpha '\pi}{L\over \sqrt{G}}$. In these variables the
the backgrounds take the form:
$$G_{IJ}(X)dX^IdX^J = dX^{+}dX^{-}$$
$$\Phi(X) = -{\ell \over \alpha '}X^{+}$$
$$T(X)=
-{2\ell^2 \over L^2}\sqrt{{\ell \over X^{+}}}e^{-{X^{-}\over \ell}}
\eqn\2$$

The theory must contain the symmetry
$\hat{g}_{ab}\rightarrow e^{2\omega}\hat{g}_{ab}$ while
$\rho \rightarrow \rho- \omega$. At the quantum level this amounts
to the vanishing of the beta functions in target space. These
beta functions are given by \refmark{\betat, \banks}
$$\beta^G_{IJ} = R_{IJ} +2\nabla_I\nabla_J\Phi
- {\sigma \over 2}\partial_I T \partial_J T$$
$$\beta^{\Phi} = -{16 \over \alpha '} +4(\nabla\Phi)^2-4\nabla^2\Phi-R
-{2\sigma \over \alpha '}T^2 +{\sigma \over 2}(\nabla T)^2$$
$$\beta^T = \nabla^2T -2G^{IJ}\partial_IT\partial_J\Phi +{4\over \alpha '}T$$
The coefficient $\sigma$ was determined by Das and Sathiapalan \refmark{\betat}
to
be $\sigma = {1\over 2} {a^4\over \alpha'^2}$ where $a$ is a lattice
spacing on the world sheet used as a regulator. The factor $\sigma$
can be absorbed into the definition of $T$. These beta functions
do not vanish for the background given by (1). If one shifts the
dilaton field to $\Phi(X) = -{2\over \sqrt{\alpha '}}X^1$ and fixes
$G = 2L^2= 2\alpha '$ and the beta functions do vanish. So we
arrive at the following action describing 3+1 to 1+1 dimensional
reduction on $S^2$.
$$ S =  \int d^2x \sqrt{-\hat{g}}(
{1\over 4}e^{-2 \phi}
(\hat{g}^{ab}\partial_a\phi \partial_b\phi -
2\hat{g}^{ab}\partial_a\phi\partial_b\rho) +
\hat{R}{1\over 4}\sqrt{1\over 2\pi}(e^{-2\phi} +\phi -2\rho)
              + {1\over 2G}e^{2\rho}
) .       \eqn\3  $$

For a general sigma model the $\hat{T}_{00}$ and $\hat{T}_{01}$
components of the stress energy tensor are given by
$$4\pi \alpha '\hat{T}_{00} =
G_{IJ} (\dot{X}^I\dot{X}^J+ {X '}^I{X'}^J) - \alpha '2\Phi ''+T$$
$$4\pi \alpha '\hat{T}_{01} =
G_{IJ}(\dot{X}^I {X'}^J+{X'}^I\dot{X}^J)-\alpha '2\dot{\Phi} '$$
In terms of the action (3) these constraints become
$$
\hat{T}_{00}= {L^2\over 2G}(-\dot{\phi}^2-{\phi '}^2 +
2\dot{\phi}\dot{\rho} + 2\phi '\rho ')e^{-2\phi}
+ {1\over 2}\sqrt{1\over 2\pi}(e^{-2\phi} +\phi -2\rho)'' -
{1\over 2G}e^{2\rho}$$
$$
\hat{T}_{01}= {L^2\over G}(-\dot{\phi}\phi ' +
\dot{\phi}\rho ' + \phi '\dot{\rho})e^{-2\phi}
+ {1\over 2}\sqrt{{1\over 2\pi}}(e^{-2\phi} +\phi -2\rho \dot{)} '$$

In the back hole interior $ r < 2MG$ the compnent $g_{tt}$ becomes positive
and $\partial_t$ becomes a spacelike killing vector of the
schwarzschild solution \refmark{\doro}. The minisuperspace approximation takes
the independence of $t$ as a property of the quantum solution
and the $\hat{T}_{00}$ becomes (where dot refers to the derivative with
respect to $r^{*}$ satisfying ${dr\over dr^{*}}={2MG\over r}-1$,
which is a timelike coordinate in the blackhole interior) :
$$
{1\over 4}(-\dot{q}^{2}+\dot{\rho}^{2})e^{2(q-\rho )} -{1\over 2G}e^{2\rho
}=0$$
Where $q=\rho-\phi$. In terms of canonical momentum
$\pi_q= -{\dot{q} \over 2}e^{2(q-\rho )}v$
$\pi_{\rho}= {\dot{\rho} \over 2}e^{2(q-\rho )}v$
and taken
as a quantum equation we obtain the Wheeler-DeWitt equation
$$
(\pi_q^2 -\pi_{\rho}^2 + {v^2\over 2G}e^{2q})\psi (q,\rho )= 0\eqn\4$$
This Wheeler-DeWitt equation has two basic solutions
$$\psi_1 (x,\rho) = e^{ik\rho }K_{i|k|}({x \over \sqrt{2G} })$$
$$\psi_2 (x,\rho) = e^{ik\rho }I_{-i|k|}({x \over \sqrt{2G} })$$
where $x= {2v \over \sqrt{2G}} Le^q$ and $K_\nu(z),I_\nu(z)$ are
modified Bessel functions. The former decays
exponentially for large $x$ and is real. It yields
a superposition of plane waves for small $x$. The latter exponentially
grows for large $x$ and is complex. It yields a single plane wave
for small $x$.

The universal Green function is the path integral from one spacelike slice
to another, and has dependence on the initial and final values of
the two dimensional fields.
$$G(x',\rho '; x, \rho) =
\int_{q=\log{x/v},\rho=\rho}^{q=\log{x'/v},\rho=\rho '}
DqD\rho e^{i\int
d\tau( {1\over 4}(-\dot{q}^{2}+\dot{\rho}^{2})e^{2(q-\rho)} +{1\over
2G}e^{2\rho})}$$
It can be evaluated by the method
of chitre and Hartle and yields the expression
$$G(x',\rho '; x, \rho) = -{1\over 2\pi} \int_{-\infty}^{\infty}
dk e^{ik(\phi '-\phi)}
K_{i|k|}({x_{>} \over \sqrt{2G} })I_{i|k|}({x_{<} \over \sqrt{2G} })$$
where $x_{>}$ or $x_{<}$ is the greater or lesser of $x,x'$. The
product structure is consistent with a general result of Brown and Martinez
\refmark{\bro}. Note that
this is identical with the loop-loop transition amplitude
of
$c=1$ string theory \refmark{\moor}.

If one introduces a quantum field
$$\Psi (x,\rho ) = \sum_k( A^1_k \psi^1_k + A_k^{1\dag} \psi_k^{1*})
= \sum_k( A^2_k \psi^2_k + A_k^{2\dag} \psi_k^{2*})$$
then the creation operators $A_k^{1\dag},A_k^{2\dag}$ act in
an extended Hilbert space. It can be verfied, using the classical solutions'
that the parameter $k$
which is the eigenvalue of the canonical momentum $\pi_{\rho}$ is related
to the blackhole mass by $k= -2vM$. Thus the above expansion can be
interpreted a sum over all black holes of variable mass.
The universal Greens function can then be represented as:
$$G(x',\rho '; x, \rho) = <0^1|T\Psi(x',\rho ') \Psi(x,\rho )|0^2>$$
Where $T$ represents ordering with respect to $x$. As the $\psi_k$
themselves are ground states of a second quantized theory
the introduction of $|0^1>$ and $|0^2>$ as states annihilated by $A_k^1$
and $A_k^2$ repectively this represents in effect a third
quantized description of the path integral. This third quantized
description of the path integral at first sight does not appear
to offer anything new but as we shall see the non-uniqueness of the
third-quantized vacuum has importnant implications for the nonconservation
of universe number and topology change for this $2d$ dilaton gravity.

It is actually necessary to inspect the universal Greens function to decide
whether third quantization is appropriate. For example Chamseddine's
description of the 2d dilaton gravity \refmark{\cham}
$S= \int d^2x \Omega(x) \sqrt{-g} R$ resulting
from the dimenensional reduction of $2+1$ dimensional gravity, yields
an entirely different type of universal Greens function. Essentially
the  $\rho$ field is entirely nailed down to its classical value inside
the path integral. Similar statements also apply to the Chern-Simon
description of $2+1$ dimensional
gravity \refmark{\carl}. There the path integral for the action
$\int e\wedge R$ reduces to a sum over flat classical solutions
\refmark{\witt}.

Now we can apply this formalism of 2d dilaton gravity to the dynamics
of a black hole interior. It has been known since the work of
DeWitt \refmark{\dewi} that the classical solutions to the
Einstein equations are given by geodesics in the space of metrics. For the
solution corresponding to the interior of the black hole the geodesic through
target space is given by:
$$e^{2\rho}={2GM\over r}-1$$
$$e^{-2\phi} = {2r^2 \over G}$$
So that upon solving the above equation for $r$ in terms of $\rho$ we
find that $q=\rho-\phi$ can be expressed as:
$$e^q = \sqrt{2M^2G}{1\over \cosh{\rho}}$$
Note that this is the same trajectory of a particle in Rindler space
\refmark{\birr}
(see figure 1).
indeed one can define new regions of target space through
$$I: \matrix{X = e^q \cosh{\rho}\cr
             T=e^q \sinh{\rho}} $$
Then one sees that as one goes from the horizon to the singularity
the original target space variables $q$ and $\rho$ only parametrize
the region of target space $X< |T|$. One can use new target space coordinates
to describe
the trajectory beyond the singularity. If we denote the region $|X|>T$ by
region II one can parametrize this region by
$$II: \matrix{X = e^{q'} \sinh{\rho'}\cr
              T = e^{q'} \cosh{\rho}}$$
The easiest way to continue the blackhole solution beyond the singularity
is to continue the solution $X= \sqrt{2M^2G} = const$ into region
II. That is to continue the upward vertical line in figure 2 into this region.

It is intriguing that the equation (4) describing the blackhole
interior is identical to a particle bouncing off an exponential
barrier i.e. the Liouvile equation. It is possible upon hitting
the barrier for a single universe to split and for a multiple
universe state to bounce back, this has
verified explicitely in two dimensional models of quantum gravity.
Similar phenomena occur for the two dimensional dilaton gravity
considered here.
In terms of the variables $X$ and $T$ the canonical momentum are defined
by $P_{X} = -{1\over 2} e^{-2\rho} \dot{X} v$ and
   $P_{T} = {1\over 2} e^{-2\rho} \dot{T} v$. The Wheeler-DeWitt
equation becomes :
$$(-P_T^2 +P_X^2 + {v^2 \over 2G})\Psi (X,T)= 0$$
The solution is simply
$$\psi_K^{out} = e^{iKX-i \sqrt{ K^2 + {v^2\over 2g}}T}$$
Promoting this solution to a field one has the expansion
$$\Psi (x,\rho ) = \sum_K( A_K \psi_K^{out} + A_K\dag \psi_K^{*out})$$
Third quantization interprets $A_K$ and $A_k^1$ as operators and allows
one to define a no-universe state or third quantized vacuum by
$A_K|out> =0$ and $A_k^1|in>=0$. The interesting fact here is that
in expanding one solution in terms of the other one finds
that the (out) vacuum is not annihilated by the $A_k^1$ in particular
$$<out| A_k^{1\dag} A_k^1 |out> = { 1\over {e^{2\pi |k|} -1}}$$
The computation is essentially the same as counting the number of
Rindler particles in Minkowski space \refmark{\full}. Here the interpretation
is that
the extended Hilbert space of third quantization has allowed the
universe number to fluctuate from the initial one universe state
of the black hole interior emerging at the horizon.

So far all of our discussion has been based the two dimensional
dilaton gravity described by (1). What possible connection does this
have with the real four dimensional world?  Perhap the hint
of such a connection lies in the work of Verlinde and Verlinde \refmark{\verl}.
In this work they studied 4d gravity in the gauge
$$ [g_{\mu \nu}] = \pmatrix {g_{ab} & 0\cr 0 & h_{ij} }$$
where the action can be written \refmark{\verl}:
$$
\eqalign{
S= \int \sqrt{-g} (\sqrt{h} R_h + {1 \over 4}&\sqrt{h}h^{ij}
\partial_i g_{ab} \partial_j g_{cd} (g^{ab}g^{cd}-g^{ad}g^{bc})) \cr
+\sqrt{h}(\sqrt{-g} R_g + {1 \over 4}&\sqrt{-g}g^{ab}
\partial_a h_{ij} \partial_b h_{kl} (h^{ij}h^{kl}-h^{il}h^{jk}))\cr}
$$
This action still contains the residual gauge invariance
$g_{ab} \rightarrow g_{ab} + \nabla_a\epsilon_b +\nabla_b\epsilon_a$
if $\epsilon_a$ is independent of $x_2,x_3$. We can fix this symmetry
by choosing the conformal gauge
$g_{ab} = e^{2\rho}\hat{g}_{ab}$ and if we parametrize
$$
[h_{ij}] = L^2e^{-2\phi}\pmatrix{
{{\tau_1^2 +\tau_2^2}\over \tau_2} & {\tau_1 \over \tau_2}\cr
{\tau_1 \over \tau_2}              & {1 \over \tau_2}}$$
the action becomes:
$$\eqalign{
 S = {L^2 \over 4 G }
          \sum_{\ell_2\ell_3}\int d^2x \sqrt{-\hat{g}} ( e^{-2 \phi}(
2\hat{g}^{ab}\partial_a\phi \partial_b\phi -
4&\hat{g}^{ab}\partial_a\phi\partial_b\rho +
\hat{R}
              + \chi L^{-2}e^{2 \phi+2\rho})\cr
 - {e^{-2\phi }\over 2\tau_2^2} \hat{g}^{ab}(\partial_a\tau_1\partial_b\tau_1+
\partial_a\tau_2\partial_b\tau_2)
+ {2e^{2\rho} \over L^2\tau_2} &( (\Delta_2\rho)^2
-2 \tau_1\Delta_2\rho\Delta_3\rho +
(\tau_1^2 +\tau_2^2)\Delta_3\rho\Delta_3\rho ) )
  \cr}  $$
Where $\chi={1\over 4\pi}\int \sqrt{h}R_h$ is the Euler characteristic of
the transverse space, which if taken to be toplogically a sphere
equals 2. The variables $\tau_1,\tau_2$ encode the two remaining degrees
of freedom of the graviton. $\Delta_i\rho = a^{-1} (\rho(x_i + a)- \rho(x_i))$
is the
finite difference defined on a lattice with lattice spacing $a$. Indeed
a similar $1+1$ dimensional transverse lattice approach has yielded
a tractible description of 4d QCD \refmark{\bard, \griff}.

It is possible that a Lagrangian in this form can be applied to the
blackhole entropy problem. A microcanonical description of
blackhole entropy is embodied in the formula :
$$tr(\delta(M-\hat{M})) = e^{4\pi GM^2}$$
Here the trace is over the Hilbert space of the blackhole interior
and $\hat{M}$ is a quantum mass operator which reduces to the usual
ADM value for the mass in the classical limit. However the right states
to be included in the trace as well a suitable
choice of the mass operator remains an open problem.

In paper
we have seen paper we have seen that the canonical momentum
of the $\rho$ field has eigenvalues which classically reduce to the mass.
Also the Bondi mass
$M_{Bondi} = {L^3\over 2G} e^{-2\rho - 3\phi}({\dot{\phi}}^2- {\phi'}^2)
+{L\over 2G}e^{-\phi}$, taken as a quantum operator has eigenvalues
proportional to the classical ADM mass. This gives a spacetime dependent
mass, crucial for the phenomena of mass inflation \refmark{\pois}.
However it is not clear
either is the correct expression to insert in the entropy formula.

The introduction
of a transverse lattice and working in the gauge of Verlinde and Verlinde
has reduced the Einstein-Hilbert action to type of string theory,
albeit one of a rather complicated form, so it worth noting that the entropy
of string theory is determined by:
$$tr(\delta(E-\hat{H})) = E^{-a} e^{bE}$$
In that case the Hilbert space was over multi-string states with the
dominant contribution comprising many strings in their ground state
and one highly excited string with the bulk of the energy \refmark{\frau}.
The analog
for the blackhole entropy is clearly a sum over multi-universe states
with many of the universes nearly empty and one containing nearly
all the mass $M$, but the role of these multiuniverse states in the
trace could be more subtle. Specific models of quantum
cosmology have been investigated in which the average
number of universes in the final state is given by $e^{{3\pi \over \Lambda G}}$
where $\Lambda $ is the cosmological constant. This number is so
huge that that it was given the name googelplexus \refmark{\fisc}. In the
Schwarzschild-DeSitter universe with blackhole mass $GM={1\over
\sqrt{\Lambda}}$
this number is $e^{{2\pi \over \Lambda G}}$ \refmark{\gibb}
and is equally large. Significantly this coincides with the
Euclidean saddle point computation of blackhole entropy
of the Schwarzschild-DeSitter solution.
Even if each universe carried off one bit of information the huge
number of them could contain the entropy present in the initial
state before blackhole evaporation and topology change.

\refout
\bf{Figure Captions}

figure 1. Representation of blackhole interior metric as a geodescic in target
space. Here $\rho$ going to plus or minus infinity corresponds to the
horizon and singularity respectively.

figure 2. Extended target space describing the blackhole interior. Points
$a$ and $b$ denote the horizon and singularity. The geodescic can be
continued beyond the singularity by extending the vertical line into
region II.
\bye